\documentclass[twocolumn,prl]{revtex4}
\usepackage{graphicx}
\usepackage{epsfig}
\begin{document}
\newcommand{\be} {\begin{equation}}
\newcommand{\ee} {\end{equation}}
\newcommand{\bea} {\begin{eqnarray}}
\newcommand{\eea} {\end{eqnarray}}
\newcommand{\sid} {\ensuremath{\Psi^{\dagger}}}
\newcommand{\rr} {\ensuremath{{\bf r}}}
\newcommand{\up} {\ensuremath{\uparrow}}
\newcommand{\dn} {\ensuremath{\downarrow}}
\newcommand{\no} {\ensuremath{\nonumber}}
\newcommand{\ep} {\ensuremath{\epsilon}}
\newcommand{\etal} {{\it et al.},\ }

\title{Breakdown of the Thomas-Fermi approximation for polarized Fermi gases}
\author{Rajdeep Sensarma, William Schneider, Roberto B. Diener, and Mohit Randeria}
\affiliation{Department of Physics, The Ohio State University, Columbus, OH 43210}
\begin{abstract}
We use Bogoliubov de-Gennes theory to show that the commonly used Thomas-Fermi
approximation (TFA) can fail in describing polarized unitary gases in 
anisotropic harmonic traps. We find a magnetized superfluid region inside 
the trap, with order parameter oscillations, even though there is no such 
stable bulk phase. This leads to magnetization profiles that deviate from
contours of constant potential energy. We determine how this violation scales 
with trap anisotropy and number of particles, and show that we are able to account 
for important differences between the MIT and Rice experiments. 
\end{abstract}
\maketitle

The study of strongly interacting Fermi systems using ultracold atomic gases has attracted interest across the physics community.  
The phenomena observed in these dilute gases are expected to shed light on systems as diverse as high Tc superconductors, quark-gluon plasmas and quantum chromodynamics.  The great virtue of atomic gases is the tunability of interactions using a Feshbach resonance so that the entire
BCS to BEC crossover can be studied, with the most strongly interacting, unitary regime in-between these two limits.
A series of beautiful experiments have probed condensation of fermionic pairs~\cite{JILA}, pairing gaps~\cite{Grimm}, quantized vortices~\cite{Ketterle1} and the thermodynamics~\cite{Thomas} of the crossover.  A particularly exciting new direction is the study of partially polarized gases~\cite{Ketterle2,Hulet1} with an imbalance in the number of up and down `spin' fermions.  

An important aspect of cold atom experiments is the presence of a harmonic trap.
A ``local density'' or Thomas-Fermi approximation (TFA) \cite{Pethick-Smith} has usually been adequate
to take this into account. 
The TFA asserts that the properties at point ${\bf r}$ in a `slowly varying' potential are the same as those of the uniform gas 
at a chemical potential $\mu_\sigma({\bf r}) = \mu_\sigma - V({\bf r})$ with $\sigma = \up,\dn$.
This leads to the simple result that the spatial dependence of any observable must follow contours
of constant trapping potential, which is directly testable for the densities $n_\sigma({\bf r})$.  

The two experiments on polarized, unitary Fermi gases find rather different results with respect to the TFA.
The MIT group~\cite{Ketterle2, Ketterle3}, with a large number of atoms $N = 10^7$ and a small
trap anisotropy $1/\alpha = \omega_r/\omega_z \simeq 5$, finds that the densities follow contours of $V({\bf r})$. 
On the other hand, the Rice group~\cite{Hulet2}, with smaller $N = 10^5$ and larger anisotropy $1/\alpha \simeq 50$
observes gross violations of the equipotential contour condition for the densities \cite{Mueller0}. 

Motivated by this, we have investigated the validity of the TFA using $T=0$ 
Bogoliubov-deGennes (BdG) calculations ~\cite{Torma-Machida} and scaling arguments
for the unitary gas in \emph{anisotropic}, three dimensional traps with polarization up to 40\%.
Our main results are: \hfill\break
(1) The TFA is always violated in a trap in so far as the spatial variation of the order parameter is concerned. Between
the unpolarized superfluid at the center and the fully polarized normal gas at the edges of the trap, there is an intermediate
region which is a magnetized superfluid with an FFLO-like oscillation \cite{FFLO} of the order parameter. 
We emphasize that there is no corresponding stable phase for the uniform system.
\hfill\break
(2) We find that the size of the magnetized superfluid region depends on both polarization and anisotropy and can be
much larger than $k_f^{-1}$ for large anisotropy. 
\hfill\break
(3) The violation of the equipotential contour criterion for the magnetization 
$m({\bf r}) = n_\up({\bf r}) - n_\dn({\bf r})$ increases with increasing anisotropy $1/\alpha$,
but decreases with increasing total number of particles $N = N_\up + N_\dn$.  
\hfill\break
(4) We derive a simple condition for the consistency of the TFA:
$\delta\mu/\omega_r = (N\alpha)^{1/3}f(P) \gg 1$, where
$f$ is a function of the polarization $P = \left(N_\up - N_\dn\right)/N$.
We use this $(N\alpha)^{1/3}$ scaling of $\delta\mu/\omega_r$ to get a better
understanding of the $N$ and $\alpha$ dependences of our BdG results.
\hfill\break
(5) We are thus able to account for the differences between the MIT and Rice experiments with respect to
the question of when the magnetization should or should not follow contours of constant potential at $T=0$.
\hfill\break
(6) A general implication of our results is that the TFA based on the bulk phase diagram must be used with caution
to describe a system in a trap when there are several competing phases. 
  
{\bf Bogoliubov-deGennes equations:} 
Our approach to the problem of strongly interacting, polarized Fermi gases in anisotropic traps is to solve the Bogoliubov-deGennes (BdG) equations~\cite{deGennes}.  This is the simplest approach which goes beyond the TFA and is a generalization of the BCS-Leggett
mean field theory for a spatially inhomogeneous gas.  This method has been applied to the study of vortices in the strongly interacting regime~\cite{Rajdeep vortex}, as well as the study of polarized gases in isotropic traps~\cite{Torma-Machida}. 
For a single-channel description valid for the experimentally-relevant wide resonance, the Hamiltonian density for the polarized gas is 
\bea
H(\rr) &=& \sum_\sigma \sid_{\sigma}(\rr) [H_0({\bf r})-\mu_\sigma] \Psi_{\sigma}(\rr) \nonumber \\
& &\quad -g \sid_{\up}(\rr)\sid_{\dn}(\rr)\Psi_{\dn}(\rr)\Psi_{\up}(\rr),
\label{hamiltonian}
\eea
where $H_0 = - \nabla^2 / 2m  + V(\rr)$, $m$ is the fermion mass and we set $\hbar = 1$.  The trapping potential is
$V(\rr) = {1\over 2 } m \omega_0^2 (r^2 + \alpha^2 z^2)$, where we use cylindrical coordinates $\rr = (r, \theta, z)$.  
We define the average chemical potential $\mu =(\mu_\up + \mu_\dn)/2$ and the difference as 
$2h= \delta\mu = \mu_\up-\mu_\dn$. The mean field state is found through the solution of the BdG equations \cite{bdg-technical}
\be
\left(
\begin{array}{clrr}%
 H_0(\rr)-\mu & {\hspace{0.7cm} } \Delta(\rr) \\
 \Delta^*(\rr) & -H_0(\rr) +\mu
\end{array}
\right)\left(
\begin{array}{clrr}%
u_i(\rr) \\
v_i(\rr)
\end{array}
\right)=E_i\left(
\begin{array}{clrr}%
u_i(\rr) \\
v_i(\rr)
\end{array}
\right)\label{BdG}
\ee 
together with the gap equation, polarization and total density at zero temperature given by
\bea
\Delta(\rr)&=&g\sum_{E_i>h} u_i(\rr)v_i^*(\rr),\label{gap equation} \label{gap-eqn}\\
m(\rr)&=& \sum_{0 \le E_i < h} \left ( |u_i(\rr)|^2 +   |v_i(\rr)|^2 \right ),\label{polarization equation}\\
n(\rr)&=&m(\rr) + \sum_{E_i>h}2 |v_i(\rr)|^2.\label{number equation}
\eea
These equations are solved self-consistently for $\Delta(\rr), \mu$ and $h$
using the constraints that the total number of particles is 
$N = \int d^3\rr \, n(\rr)$ and the polarization
$P = N^{-1}\int d^3\rr \, m(\rr)$.  

\begin{figure}\label{figure1}
\includegraphics[width=3in]{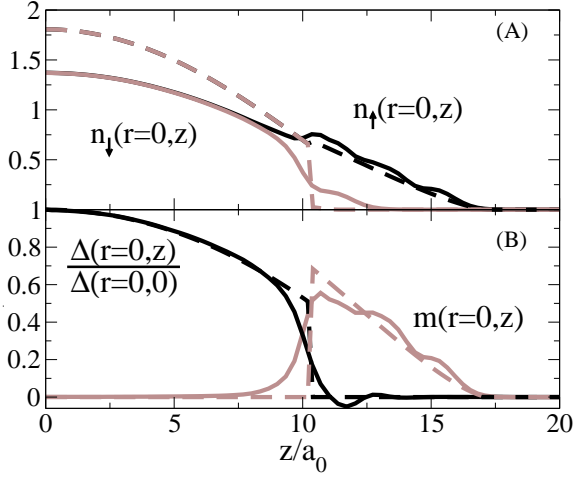}
\caption{(A) Majority (black) and minority (gray) density profiles 
and (B) order parameter $\Delta$ (black) and magnetization (grey) profiles
along the axis of the trap.  The solid lines are BdG results and the dashed lines are TFA results.
The calculations are for a trap with $\alpha = 1/4$ containing $N = 865$ particles and a polarization of $30\%$.}
\end{figure}

The solution of these equations is simplified if we expand the wavefunctions in terms of the eigenfunctions of the diagonal piece $H_0(\rr)-\mu$.  Measuring lengths in units of the \emph{radial} harmonic oscillator length $a_0 = 1/\sqrt{m\omega_0}$ and energies in units of $\omega_0$, these functions are $\phi_{np\ell} = f_{p\ell}(r) \exp(i\ell \theta) g_n(z) / \sqrt{2\pi}$,
where the radial and axial functions are related to associated Laguerre and Hermite polynomials, 
$f_{p\ell} (r) = \sqrt{{p!}/{(p+\ell) !}}e^{-r^2/2}r^{|\ell|}L^{\ell}_p(r^2)$ and $g_n(z)=
\sqrt{{\sqrt{\alpha}}/({2^n\sqrt{\pi} n!})}e^{-\alpha z^2/2}H_n(\sqrt{\alpha}z)$, respectively.
The corresponding eigenvalue is $\epsilon_{np\ell} = (2p+\ell+1) +\alpha ( n+1/2) -\mu$.  
The BdG Hamiltonian is block-diagonal in $\ell$ due to axial symmetry. For a given $\ell$ we need to diagonalize
\be 
H^{(\ell)}=\left(
\begin{array}{clrr}%
T^{(\ell)} & {\hspace{0.2cm}}\Delta^{(\ell)} \\
\Delta^{(\ell)} & -T^{(\ell)}
\end{array}
\right), 
\ee
where $T^{(\ell)}_{nn'pp'}=\epsilon_{n\ell p}\delta_{nn'}\delta_{pp'}$ and
$$
\Delta^{(\ell)}_{nn'pp'}=\int_0^{\infty} rdr\int_{-\infty}^{\infty} dz f_{p\ell}(r)f_{p'\ell}(r)g_n(z)g_{n'}(z) \Delta(r,z)
$$
Since $\Delta(r,z)=\Delta(r,-z)$, the only non-zero matrix 
elements of $\Delta_{nn'pp'}$ correspond to even $n+n'$.

The bare coupling $g$ in eqs.~(\ref{hamiltonian},\ref{gap-eqn}) is related to the two-body s-wave scattering length $a_s$
through $m/(4\pi a_s)=1/g+m^{3/2}\sqrt{E_c}/(\sqrt{2}\pi^2)$, where $E_c$ is an ultraviolet energy cutoff. 
The number of particles determines $\epsilon_f^0 = (6N\alpha)^{1/3}\omega_0$ and we have used a cutoff of 
$E_c=4\epsilon_f^0$ for the calculations in this paper.

\begin{figure*}
\includegraphics[height=2.02in]{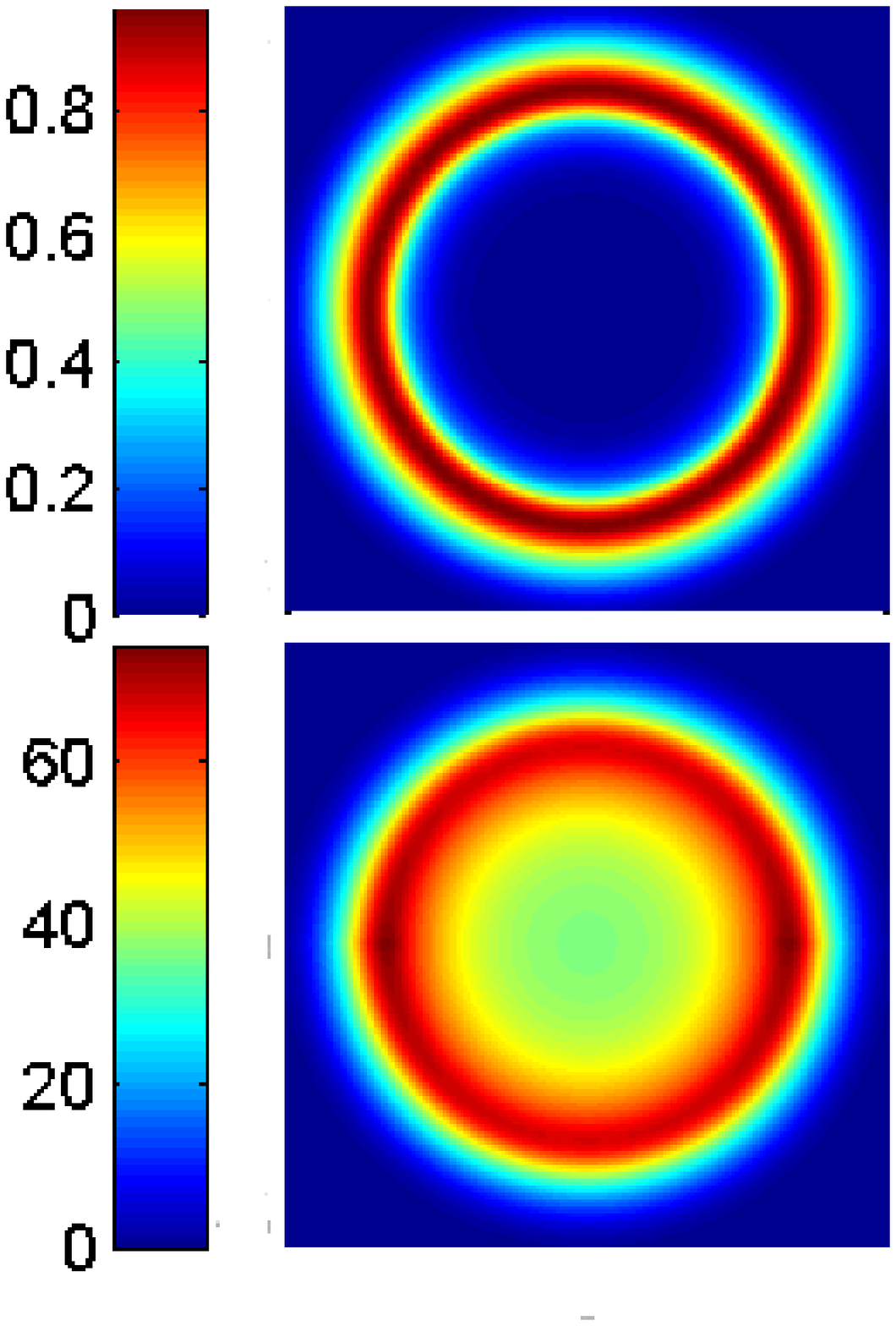}\hspace{-0.5cm}
\includegraphics[height=1.96in]{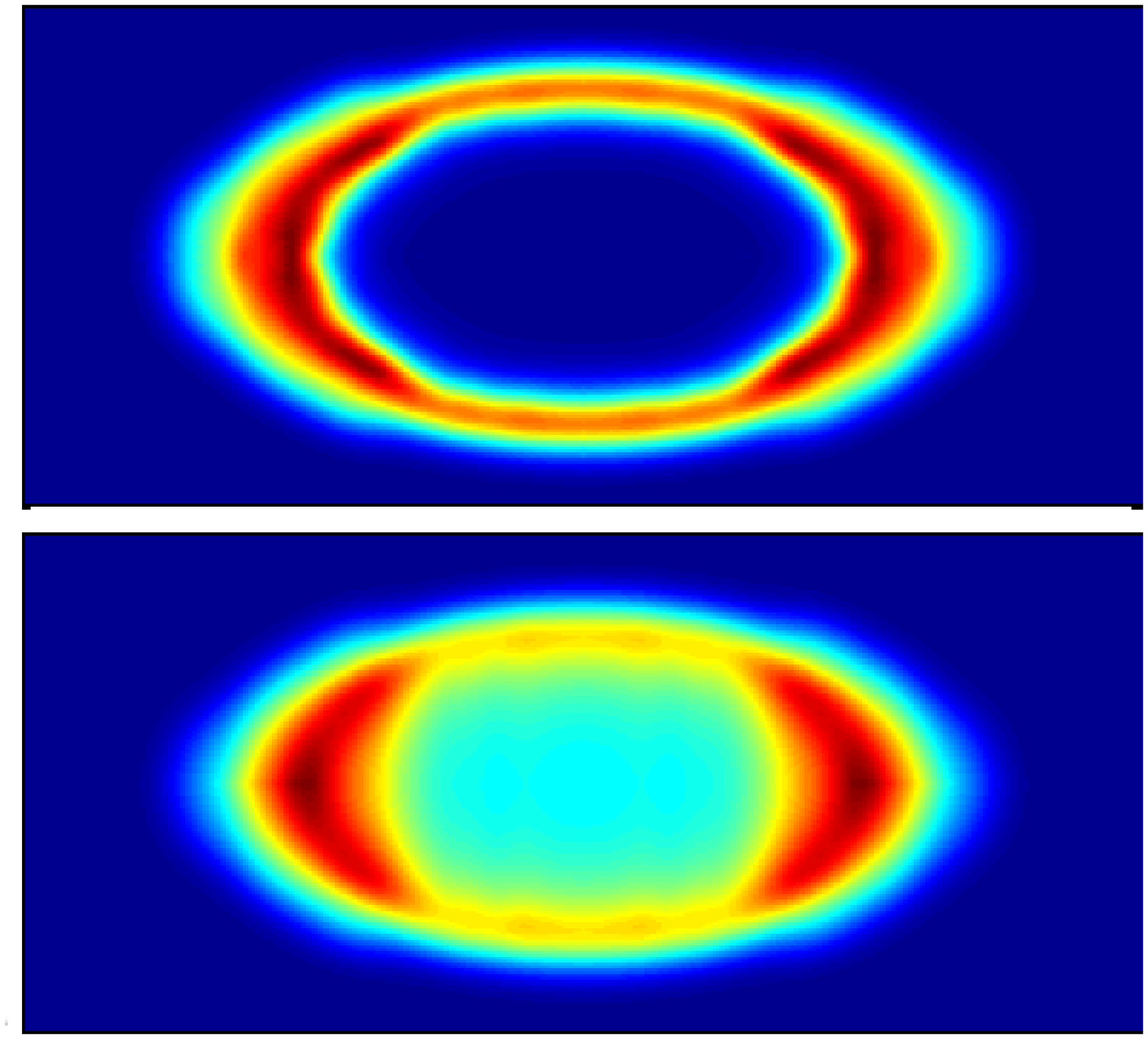}\hspace{-0.5cm}
\includegraphics[height=2.08in]{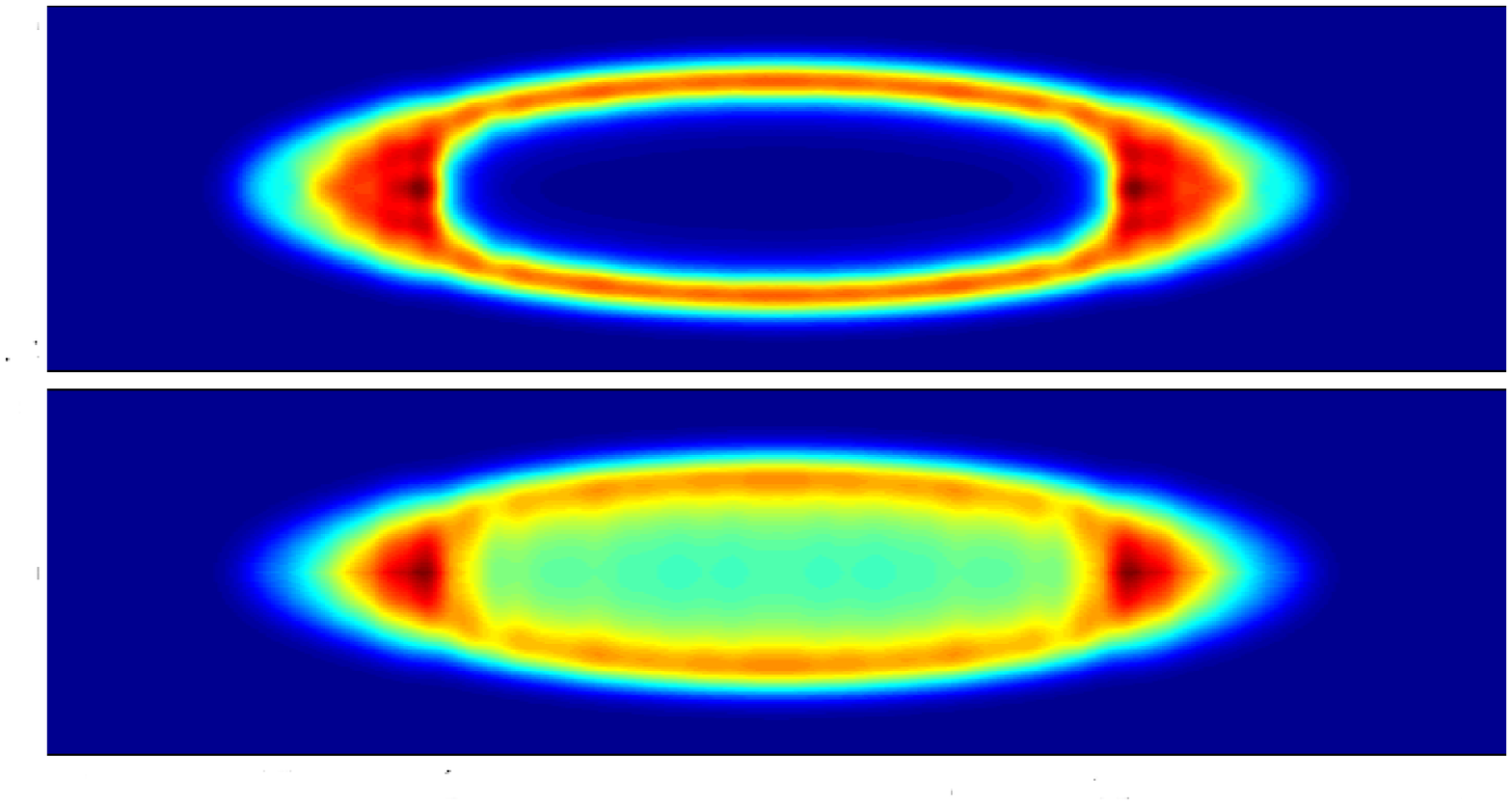}
%$\begin{array}{c@{\hspace{0.2in}}c@{\hspace{0.2in}}c}
%\epsfysize=1.0in
%\epsffile{pol-562-1-30-4.eps} &
%\epsfysize=1.0in
%\epsffile{pol-562-2-30-4.eps} &
%%\epsfxsize=1.5in
%\epsfysize=1.0in 
%\epsffile{pol-562-4-30-4.eps} \\
%\epsfysize=1.0in
%\epsffile{polint-562-1-30-4.eps} &
%\epsfysize=1.0in
%\epsffile{polint-562-2-30-4.eps} &
%%\epsfxsize=1.5in
%\epsfysize=1.0in 
%\epsffile{polint-562-4-30-4.eps}
%\end{array}$
\caption{Top row: False color plots of the 3D magnetization as a function of $(z,r)$ for a system with $N = 865$ and $P = 30\%$.  
The trapping potential has $\alpha = 1, 1/2, 1/4$ for the left, center and right panels respectively.  
Bottom row: the corresponding column integrated magnetization densities.}
\end{figure*}

{\bf Densities and Order Parameter:} 
We now discuss the self-consistent solution of the BdG equations as a function of total 
$N = N_\up + N_\dn$, polarization $P = \left(N_\up - N_\dn\right)/N$ and trap anisotropy $1/\alpha$
at unitarity ($a_s = \infty$).
We have extensively studied the problem for $N$ up to 2500 particles, $0 \le P \le 0.4$ and
$\alpha = 1, 1/2, 1/4$.   
In Fig.~1(A) we plot the majority ($n_\up$) and minority ($n_\dn$)
densities along the $z$ axis for a representative data set ($\alpha =1/4$, $N = 865$ and polarization $P=30\%$).   
In Fig.~1(B) we plot the corresponding magnetization $m({\bf r}) = n_\up - n_\dn$
together with the local order parameter $\Delta(\rr)$.
In both panels the solid lines are BdG results, while dashed lines are TFA predictions
(using the bulk phase diagram \cite{bulk phases} as input). 

Both the BdG and TFA results show an unpolarized superfluid at the center of the trap
and a fully polarized normal gas at the edge. There is a marked decrease in the
BdG central density relative to TFA, with a redistribution of minority atoms
to an intermediate region.
The main difference between BdG and TFA is precisely in this intermediate region.
Within the TFA there is a discontinuous jump in the order parameter 
which is smoothed out in the BdG solution since this lowers the gradient energy. 
What is perhaps unexpected is that the decaying BdG order parameter exhibits oscillations, 
similar to those expected in a putative FFLO phase \cite{FFLO} 
with a period roughly consistent with $2\pi / (k_{F\up} - k_{F\dn})$ (where
$k_{F\sigma}$ are the local Fermi wavevectors). 
Irrespective of the importance (or otherwise) of the limited number of rather small amplitude
oscillations, it is unambiguous that this intermediate region is a \emph{magnetized superfluid}:
it has both a non-vanishing superfluid order parameter and a non-zero magnetization.

We note that the magnetized region that we see is completely different from a partially
polarized \emph{normal} region. The latter, though barely visible, is present in the TFA results
of Fig.~1(A) and derives from the partially polarized normal phase which exists for a small range of $\mu/h$ 
in the bulk mean field phase diagram \cite{bulk phases}. Such a partially polarized normal region, 
which is also seen prominently in phenomenological implementations of the TFA \cite{Chevy},
is never observed in our BdG calculations with $P \le 40\%$.

The existence of a magnetized superfluid region implies the breakdown of TFA.
There is no such phase for a uniform gas in the thermodynamic limit at unitarity
and this region is stabilized only by the presence of a trap. 
Earlier BdG studies in isotropic traps~\cite{Torma-Machida} already found such an FFLO-like region but,
as we discuss next, it has a much larger spatial extent in the anisotropic case.  

It is important to ask if the intermediate region is sufficiently narrow that 
it can be described as an interface with a surface energy \cite{Mueller1}.  
While this may be a reasonable semi-phenomenological description, we believe it is not
microscopically correct. From our BdG calculations we find that the size of the intermediate region  
is proportional to $k_F^{-1}$ but with a large proportionality constant which is $P$ and $\alpha$ dependent,
and increases rapidly with trap anisotropy $\alpha^{-1}$ along the axial direction.
For instance, in Fig.~2 the $z$-extent of the magnetized superfluid is 6 times, 9 times and
16 times the local $k_F^{-1}$ for $\alpha = 1, 1/2$ and $1/4$ respectively \cite{size}.

We emphasize that, despite an apparently widespread feeling to the contrary, 
the size of the intermediate region is also quite large in the experiments.  
This size can be determined from the separation between the point at which
$n_\up$ and $n_\dn$ first deviate from each other to the point where $n_\dn$ vanishes. 
Even the MIT data, which shows little evidence for breakdown
of TFA, has an intermediate region of $\simeq 10 \mu$m along the radial direction, 
a significant fraction of the unpolarized core radius; see Fig.~5(b) of ref.~\cite{Ketterle3}. 
In the Rice data the intermediate region is
$\simeq 100 \mu$m along the axial direction.
This is two-orders of magnitude larger than $k_f^{-1}$ and a
significant fraction of the superfluid core size;
see Fig.~3(b) of ref.~\cite{Hulet2}. 

{\bf Equipotential contours:}
At present there are no direct experimental probes of the magnetized superfluid region.
Thus we must look for signatures of the breakdown of TFA in the testable question of whether the densities 
follow contours of constant potential.  

We first show how the equipotential contour condition is progressively violated 
as a function of increasing trap anisotropy $1/\alpha$.
In the top panel of Fig.~2 we plot the magnetization $m(r,z)$ for $N=865$ particles with $P=30\%$ polarization
and $\alpha = 1, 1/2, 1/4$. In the lower panel we show the corresponding plots of the 
column-integrated magnetization, which is simpler to measure in experiments, and is given by 
$m_{col}(y,z) = \int_{-\infty}^\infty  dx \,m(\sqrt{x^2+y^2},z)$.
For the spherical case ($\alpha=1$) the equipotential contour condition must be satisfied by symmetry, despite the violation of TFA
in the order parameter. As the anisotropy $\alpha^{-1}$ increases we see that the magnetization gets more concentrated 
along the wings.  Moreover, the boundary between the magnetized and unmagnetized regions near the $z=0$ plane becomes straighter, yielding a magnetization ``hole" that becomes more rectangular.  This is very similar to the observed profiles in the 
Rice experiments \cite{Hulet2}.

\begin{figure}\label{figure3}
\includegraphics[height=3in]{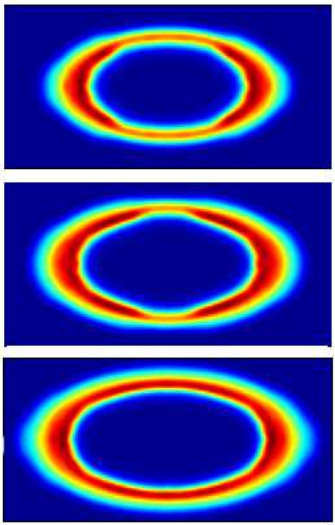}
\caption{Magnetization profiles as a function of total N (from top to bottom: 865, 1538 and 2307) at a fixed trap anisotropy ($\alpha = 1/2$) and  $P = 30\%$.}
\end{figure}

To understand why the MIT results look so different, we must study the dependence on the total number of particles,
for a fixed trap anisotropy and polarization. 
In Fig.~3 we plot the results for (from top to bottom) $N =$ 865, 1538, and 2307 particles in a trap with $\alpha = 1/2$ and $P= 30\%$.  
For the smallest $N$ the magnetization is localized along the axis, and is 
seen to spread out toward the radial direction with increasing $N$. The largest $N$ results show a rather
elliptical magnetization density indicating that the magnetization begins to follow the equipotential contours as $N$ increases.

{\bf Scaling with N and $\alpha$:} 
We next present a simple argument for the consistency of the TFA which will
allow us to see better how our results scale with $N$ and $\alpha$.
In a polarized Fermi gas the conditions for the consistency of the TFA are $\mu_\sigma/\omega_0 \gg 1$
and $\delta\mu / \omega_0 \gg 1$. The last inequality is the one that is most easily violated when
$P$ is not close to unity. We start with the TFA densities obtained from the mean-field 
phase diagram at unitarity \cite{bulk phases}. By spatially integrating these density profiles
we find $N_\up$ and $N_\dn$ in terms of $\delta\mu$ and the average $\mu$.
Inverting these relations we can show that $\delta \mu / \omega_0 = (N \alpha)^{1/3} f(P)$
\cite{Mueller3}. Here $f(P)$ is a monotonically increasing function of $P$ which goes
like $P^{2/5}$ for $P \ll 1$ and is of order unity for the $P$ values of interest. 

We have checked that our BdG results for $\delta \mu / \omega_0$  are consistent with the
$(N \alpha)^{1/3}$ scaling, even though the values are smaller than the TFA estimates.
If the condition $\delta \mu / \omega_0 \gg 1$ is violated the TFA
will breakdown. How this breakdown manifests itself in the size of the magnetized superfluid and in
the violation of the equipotential contour criterion depends strongly on the anisotropy $1/\alpha$,
as seen above. For a given $(N \alpha)^{1/3}$ these violations are more pronounced for larger $1/\alpha$.

The Rice experiments with $(N\alpha)^{1/3} \simeq 10$ and a large anisotropy
show a significant violation of the equipotential contour criterion, consistent with
our findings. On the other hand, we also understand why the MIT experiments, 
with $(N\alpha)^{1/3} \simeq 100$ and a small anisotropy, find that
this criterion is obeyed.

In summary, we have shown using BdG equations for the unitary, polarized Fermi gas
that the often used TFA breaks down with the appearance of an intermediate magnetized
superfluid region. The deviations from TFA are more pronounced with increasing trap 
asymmetry, both in terms of the size of the intermediate region and in the 
violations of the equipotential contour criterion.
These violations become less important for large particle numbers and more spherical traps, 
in a way that is consistent with current experimental results.
Important questions for future work are inclusion of finite temperature effects, 
ways to probe the intermediate magnetized superfluid region,
and understanding the nature of the ground state at large polarization. 

The authors would like to acknowledge the use of facilities of the Ohio Supercomputing Center, and thank
Jason Ho for numerous conversations and Brian Peters for help with the preparation of the figures.


\begin{thebibliography}{99}
\bibitem{bdg-technical}
The BdG Hamiltonian with $\mu_\up \ne \mu_\dn$ 
can be written as $H_{\rm BdG} = H^0_{\rm BdG} - h I$ where 
$H^0_{\rm BdG}$ is the operator in eq.~(\ref{BdG}) 
and $I$ the unit operator. Thus diagonalizing $H_{\rm BdG}$ is equivalent to 
eq.~(\ref{BdG}) in which only $\mu$
enters, with $h$ appearing in the self-consistency conditions.
\bibitem{size}
We can see the factor of 16 for $\alpha = 1/4$
in Fig.~1, where the z-extent of the intermediate region $\simeq 4 a_0$ 
while the local $k_F^{-1} = [3\pi^2n(\rr)]^{1/3} \simeq a_0/4$.
\bibitem{Mueller3}
A similar criterion (without an explicit $P$-dependence)
for the violation of the equipotential-contour condition was derived in 
ref.~\cite{Mueller1} using a different approach.
\end{thebibliography}
\end{document}